\documentclass{article}

\usepackage{microtype}
\usepackage{graphicx}
\usepackage{subfigure}
\usepackage{booktabs} 

\usepackage{hyperref}

\usepackage[accepted]{icml2021}

\icmltitlerunning{On Perceptual Lossy Compression: The Cost of Perceptual Reconstruction and An Optimal Training Framework}

\usepackage{amsmath}
\usepackage{amssymb}
\begin{document}

\twocolumn[
\icmltitle{On Perceptual Lossy Compression: The Cost of Perceptual\\Reconstruction and An Optimal Training Framework}



\icmlsetsymbol{equal}{*}

\begin{icmlauthorlist}
\icmlauthor{Zeyu Yan}{edu}
\icmlauthor{Fei Wen}{edu}
\icmlauthor{Rendong Ying}{edu}
\icmlauthor{Chao Ma}{edu}
\icmlauthor{Peilin Liu}{edu}

\end{icmlauthorlist}

\icmlaffiliation{edu}{School of Electronic Information and Electrical Engnieering, Shanghai Jiao Tong University, Shanghai, China}

\icmlcorrespondingauthor{Fei Wen}{wenfei@sjtu.edu.cn}


\vskip 0.3in
]



\printAffiliationsAndNotice{}  

\begin{abstract}
Lossy compression algorithms are typically designed to achieve the lowest possible distortion at a given bit rate. However, recent studies show that pursuing high perceptual quality would lead to increase of the lowest achievable distortion (e.g., MSE). This paper provides nontrivial results theoretically revealing that, \textit{1}) the cost of achieving perfect perception quality is exactly a doubling of the lowest achievable MSE distortion, \textit{2}) an optimal encoder for the “classic” rate-distortion problem is also optimal for the perceptual compression problem, \textit{3}) distortion loss is unnecessary for training a perceptual decoder. Further, we propose a novel training framework to achieve the lowest MSE distortion under perfect perception constraint at a given bit rate. This framework uses a GAN with discriminator conditioned on an MSE-optimized encoder, which is superior over the traditional framework using distortion plus adversarial loss. Experiments are provided to verify the theoretical finding and demonstrate the superiority of the proposed training framework.

\end{abstract}

\section{Introduction}
\label{submission}

Lossy compression is a fundamental problem in modern digital world for efficient transmission and storage of image, video and audio data. Recently, due in part to the roaring success of deep learning, the research of deep neural networks (DNNs) based compression has attracted much attention and shown promising results in image, audio, and video compression \cite{2017RealTime, 2018DeepGenerative, 2019Extreme, 2016Variable, 2017FullResolution, 2016End2end, 2017End2end, 2018Variational, 2017Soft2hard, 2018JointAutoregressive, 2018Learning, 2018Conditional, 2018Improved, 2017DeepGenerative, 2018Generative, 2018Deformation}. For lossy compression, Shannon’s rate-distortion theory is a theoretical cornerstone, which characterizes the tradeoff between the bit rate of compressed representation and the distortion in reconstructing the data \cite{1959Coding, ThomasCover2nd}.

Typically, lossy compression algorithms are designed based on the rate-distortion theory to achieve the lowest possible distortion at a given bit rate. In this context, lower distortion is desired, e.g., lower mean square error (MSE), higher peak signal to noise ratio (PSNR), or higher structural similarity (SSIM) \cite{2004ImageQuality}. However, recent studies have demonstrated that these distortion measures are not fully consistent with human’s perception \cite{2016Perceptual, 2018Unreasonable, 2018PDtradeoff, 2019Rethinking, 2019Extreme, 2017DeepGenerative, 2018Generative, 2017LearningAnoreference}. Specifically, minimizing distortion alone does not necessarily lead to good perceptual quality. In fact, it has been shown that pursuing high perceptual quality would lead to increase of the lowest achievable distortion (e.g., MSE). Generally, there exist two methods to improve perceptual quality. The first is to incorporate a perceptual loss measuring the difference between deep features \cite{2016Perceptual, 2015VeryDeep, 2016ImageStyle, 2017Photographic, 2016Generating}. The second is to incorporate an adversarial loss by using generative adversarial networks (GAN) \cite{2014GAN}. Noteworthily, using an adversarial loss, the second method has shown remarkable effectiveness in achieving high perceptual quality \cite{2017RealTime, 2019Extreme, 2017PhotoRealistic, 2019ESRGAN, 2020AGANbased, 2020Fidelity, 2020HighFidelity}.

For perceptual reconstruction, new insights from recent studies have revealed that distortion and perceptual quality are at odds with each other. It can be well characterized by a perception-distortion tradeoff \cite{2018PDtradeoff}, in which perceptual quality is defined in terms of the deviation between the distributions of the source and the reconstructed data. More recently, this result has been extended to the lossy compression problem, resulting in a rate-distortion-perception tradeoff \cite{2018IntroducingPD2RD, 2018RDPtradeoff, 2019Rethinking}. The three-way tradeoff indicates that imposing a high perceptual quality constraint on the lossy compression problem would lead to an elevation of the rate-distortion curve. Hence, a sacrifice in either rate or distortion is necessary to achieve high perceptual quality. The works \cite{2018PDtradeoff, 2019Rethinking} have paved the way for understanding the perception-distortion tradeoff.

While it has become increasingly accepted that high perceptual quality can be achieved with some increase of the lowest achievable distortion, there lacks quantitative analysis on such an increase of distortion. This work is motivated by the following two open questions:

1. Is it possible to quantitatively characterize the effect of\\
\hspace*{0.35cm}perceptual constraint on the rate-distortion tradeoff?\\
2. How to build a framework for perfect perception\\
\hspace*{0.35cm}reconstruction in lossy compression?

Toward answering these two questions, the main contributions of this work are as follows.

First, we derive a nontrivial result theoretically revealing that, at a given bit rate, the cost of achieving perfect perceptual quality is exactly a doubling of the lowest achievable MSE distortion. Our analysis also shows that an encoder optimized in terms of MSE is also an optimal encoder under perfect perception constraint. This result implies that the commonly used adversarial loss in state-of-the-art works \cite{2019Rethinking, 2019Extreme, 2020Fidelity} is in fact unnecessary for optimizing the encoder.

Second, based on the analysis, we propose a training framework that can achieve the lowest MSE distortion under perfect perception constraint at a given bit rate. This framework uses a GAN with its discriminator conditioned on an encoder optimized in terms of MSE, which is superior over the traditional distortion plus adversarial loss (DAL) based framework.

Finally, experimental results on the MNIST dataset are provided to verify the theoretical finding and demonstrate the superiority of the proposed training framework. Particularly, extensive experimental results show that the effect of perception constraint on the rate-distortion tradeoff accords well with our theoretical result, i.e., a doubling of the MSE distortion in achieving high perceptual quality.

Though our result is restricted to the lossy compression problem, to the best of our knowledge, it is the first quantitative result on the lowest degradation of distortion in achieving perceptual reconstruction. Note that the work \cite{2019Rethinking} has shown that perfect perceptual quality can be attained at a sacrifice of no more than 2-fold increase in MSE distortion. Our analysis is fundamentally different from \cite{2019Rethinking} in three aspects. \textit{1}) We obtain a different result deterministically shows that the increase is exactly 2-fold. \textit{2}) The result in \cite{2019Rethinking} is derived through analyzing a constructed encoder-decoder pair that achieves perfect perceptual quality by concatenating a post-processing perceptual mapping. In contrast, our analysis follows a different line via directly analyzing the perception constrained lossy compression formulation itself. \textit{3}) Our analysis also provides a new insight that, to achieve high perceptual quality, perceptual loss is unnecessary for training an encoder. Besides, distortion loss, either on pixels or deep features, is unnecessary for training a decoder. We show that an adversarial loss, with the discriminator conditioned on an encoder optimized in terms of MSE distortion, is enough to achieve perfect perceptual quality. A more detailed comparison is provided in Section 4.2.

\section{Preliminaries}

\subsection{Perceptual Quality}

In image processing, peak signal to noise rate (PSNR), MSE and SSIM/MS-SSIM \cite{2003Multiscale, 2004ImageQuality} are commonly used distortion measures. However, recent studies have shown that these measures are not fully consistent with human’s perception \cite{2016Perceptual, 2018Unreasonable, 2019Extreme, 2017DeepGenerative, 2018Generative}. An image with lower distortion does not necessarily have better perceptual quality. In practice, it has been empirically shown that high perceptual property can be achieved at the cost of increased distortion. This behavior can be theoretically characterized as a perception-distortion tradeoff, which has been put forward in \cite{2018PDtradeoff}. It reveals that minimizing distortion would cause the distribution of reconstructed outputs deviating from that of the (ground-truth) source, which leads to worse perceptual quality.

The perceptual quality of a restored sample is the extent to which it looks like a natural sample from human’s perception, regardless its similarity to any reference sample. It can be conveniently defined in terms of the deviation of the distribution of restored outputs $\hat{X}$ from that of natural samples $X$ as \cite{2018PDtradeoff}
\begin{align}
d({p_X},{p_{\hat X}}),
\end{align}
which is some divergence (e.g., the Kullback-Leibler divergence or Wasserstein distance) satisfying $d(p,q) \ge 0$ and
$d(p,q) = 0 \Leftrightarrow p = q$ for any distributions $p$ and $q$. Such a definition conforms with the common practice of quantifying perceptual quality through real-versus-fake questionnaire studies \cite{2018Unreasonable, 2016Colorful, 2016ImprovedTech}. Basically, this definition of perceptual quality is based on the deviation from natural sample statistics, which correlates well with human subjective score and has been widely used in designing no-reference image quality measures.\cite{2013Making, 2005Reduced}

\subsection{Rate-Distortion-Perception Tradeoff}

Shannon’s rate-distortion theory characterizes the fundamental tradeoff between the rate (bits per symbol) and the expected distortion. Specifically, the relation between the input $X\sim{p_X}$ of the encoder and the output $\hat X\sim{p_{\hat X}}$ of the decoder can be viewed as a mapping defined by a conditional distribution $p_{{\hat X}|X}$. The information rate-distortion function is defined as \cite{ThomasCover2nd}
\begin{align}
\begin{split}
&{R^{(I)}}(D) = \mathop {\min }\limits_{{p_{\hat X|X}}} I(X;\hat X)\\&\text{s.t.}\quad \mathbb{E}{\rm{[}}\Delta {\rm{(}}X,\hat X{\rm{)]}} \le D, \label{RDfunction}
\end{split}
\end{align}
where $I$ stands for mutual information, and $\Delta$ is a distortion measure such as MSE or hamming distance.
It has been well established that ${R^{(I)}}(D)$ is a convex and non-increasing function of $D$, which demonstrates the rate-distortion tradeoff.

The rate-distortion function \eqref{RDfunction} does not take into account the perceptual quality of reconstruction. To take the perceptual quality into consideration, the traditional rate-distortion tradeoff has been extended to a three-way tradeoff model including rate, distortion, and perception as \cite{2019Rethinking}
\begin{align}
\begin{split}
&{R^{(I)}}(D,P) = \mathop {\min }\limits_{{p_{\hat X|X}}} I(X;\hat X)\\&\text{s.t.}\quad \mathbb{E}{\rm{[}}\Delta {\rm{(}}X,\hat X{\rm{)]}} \le D,\ d({p_X},{p_{\hat X}}) \le P, \label{IRDPfunction}
\end{split}
\end{align}
where $d({p_X},{p_{\hat X}})$ is a divergence between the distributions of the source and the reconstruction output, typically measured by Kullback-Leibler divergence or Jensen-Shannon divergence.
When $P =  + \infty $, the perception constraint is invalid and \eqref{IRDPfunction} degenerates to the traditional rate-distortion function \eqref{RDfunction}. When $P =  0 $, the distributions of $X$ and $\hat X$ are constrained to be identical, which defines a rate-distortion function under perfect perceptual quality constraint. It has been shown in the work \cite{2019Rethinking} that, the rate-distortion curve necessarily elevates when imposing the perfect perceptual quality constraint, but the elevation is bounded. Specifically, for the MSE distortion, the rate-distortion function under perfect perception constraint, i.e., $R^{(I)}(D,0)$, is upper bounded by a scaled version of the rate-distortion curve under no perception constraint as
\begin{align}
R^{(I)}(D,0) \le R^{(I)}\left( {\frac{1}{2}D, + \infty } \right).
\end{align}
It implies that, at a given bit rate, the increase of the MSE distortion incurred by perfect perceptual quality constraint is no more than 2-fold the MSE distortion in the case without any perception constraint. However, how much is the elevation of the rate-distortion curve incurred by the perfect perceptual quality constraint is still unclear.

\subsection{Adversarial Loss for Perceptual Reconstruction}

To improve the perceptual quality of reconstruction, a natural way is to minimize the deviation from the distribution of natural samples. As GAN is very effective in aligning distributions, the GAN-based methods have shown remarkable improvement in perceptual quality \cite{2018Unreasonable, 2019Extreme}. For the lossy compression problem, a typical formulation incorporating an adversarial loss to optimize the encoder $E$ and the decoder $G$ through adversarial training is given by
\begin{align}
\mathop {\min }\limits_{E,G} {{\cal L}_{rec}}{\rm{ + }}\lambda {{\cal L}_{adv}}{\rm{ + }}\beta H,
\end{align}
where ${\cal L}_{rec}$ is a distortion loss (e.g., MSE, $L_1$ norm, or distance between feature maps), ${\cal L}_{adv}$ is an adversarial (perception) loss which is measured by a discriminator, and $H$ is the entropy of the compressed representation. $\lambda $ and $\beta$ are positive parameters which balance the three terms. Although adversarial training helps to improve the perceptual quality, such a training framework has limitations. Specifically, if the value of $\lambda $ is not large enough, the resulting perceptual quality would not be satisfactory. On the other hand, when $\lambda $ is a relatively large value, the distortion of the system cannot be well optimized and would finally result in excessive increase in distortion, as will be shown later in experiments. Hence, it is difficult to achieve the lowest distortion under perfect perception constraint by simply balancing the distortion loss ${\cal L}_{rec}$ and the adversarial loss ${\cal L}_{adv}$.

To address these limitations, we propose a training framework that can achieve the lowest distortion under perfect perception constraint. It is also based on GAN but avoids the balance between the distortion and adversarial losses.

\section{Main Results and Proposed Training Framework}

\subsection{Analysis on the Rate-Distortion Tradeoff Under Perfect Perception Constraint}

Suppose that $X$ is a discrete source with a finite alphabet $\chi {\rm{ = \{ }}{x_i} \in {\mathbb{R}^N}:{\rm{1}} \le i \le m{\rm{\} }}$, e.g., $m = {256^{3N}}$ for 8-bit RGB image with $N$ pixels. From \eqref{IRDPfunction}, the information rate-distortion function under perfect perception constraint is
\begin{align}
\begin{split}
&{R^{(I)}}(D,P) = \mathop {\min }\limits_{{p_{\hat X|X}}} I(X;\hat X)\\&\text{s.t.}\quad \mathbb{E}{\rm{[}}\Delta {\rm{(}}X,\hat X{\rm{)]}} \le D,\ d({p_X},{p_{\hat X}}) \le 0. \label{IRDP0function}
\end{split}
\end{align}
Under the constraint that the distributions of $X$ and $\hat X$ are identical, $\chi$ is also the alphabet of $\hat X$.

Before proceeding to the analysis, we present some definitions. Define a joint distribution matrix ${\bf{B}} \in {\mathbb{R}^{m \times m}}$, of which the elements are given by
\begin{align}
{b_{ij}}: = p(X = {x_i},\hat X = {x_j}), {\rm{1}} \le i,j \le m.
\end{align}
Meanwhile, define a distortion matrix ${\bf{W}} \in {\mathbb{R}^{m \times m}}$, of which the elements are given by
\begin{align}
{w_{ij}}: = \Delta ({x_i},{x_j}), {\rm{1}} \le i,j \le m.
\end{align}
Then the distortion constraint in \eqref{IRDP0function} can be rewritten as $\left\langle {{\bf{W}},{\bf{B}}} \right\rangle  \le D$. Moreover, let ${G_{{p_X}}}({\bf{B}})$ denote the objective function of \eqref{IRDP0function}, then under perfect perception constraint, it follows that
\begin{align}
\begin{split}
{G_{{p_X}}}({\bf{B}}): &= I(X;\hat X)\\
 &= H(X) + H(\hat X) - H(X,\hat X)\\
 &= 2H(X) + \sum\limits_{i = 1}^m {\sum\limits_{j = 1}^m {{b_{ij}}} \log {b_{ij}}},
\end{split}
\end{align}
where the last equaity follows from the fact that $X$ and $\hat X$  have the same distribution under the constaint $d({p_X},{p_{\hat X}}) \le 0$. For fixed $p_X$, $p_{{\hat X}|X}$ can be equivalently represented by ${\bf{B}}$. Thus, the formulation \eqref{IRDP0function} can be rewritten as
\begin{align}
\begin{split}
{R^{(I)}}(&D,0) = \mathop {\min }\limits_{\bf{B}} {G_{{p_X}}}({\bf{B}})\\
\text{s.t.}\quad &\left\langle {{\bf{W}},{\bf{B}}} \right\rangle  \le D,\\
&\sum\limits_{i = 1}^m {{b_{ij}}} = \sum\limits_{i = 1}^m {{b_{ji}}}  = p(X = {x_j}),{\rm{1}} \le j \le m \label{IRDP0FinM}.
\end{split}
\end{align}
Note that the perception constraint $d({p_X},{p_{\hat X}}) \le 0$ in \eqref{IRDP0function}, which means ${p_X}={p_{\hat X}}$, has been rewritten as the equivalence constraint between the row summation and column summation of ${\bf{B}}$, which are the distributions of $X$ and $\hat X$, respectively. Based on \eqref{IRDP0FinM}, we have the following result (proof is given in the supplimentary material).

\textbf{Lemma 1.} \textit{Suppose that} $\Delta$ \textit{is symmetric, then any optimal solution} $\bf{B}^*$ \textit{to} \eqref{IRDP0FinM} \textit{is a symmetric matrix.}

Lemma 1 shows that when the distortion measure is symmetric (e.g., squared-error), for any optimal encoder-decoder pair to \eqref{IRDP0FinM} (equivalently \eqref{IRDP0function}), the joint distribution of $X$ and $\hat X$ is also symmetric as
\begin{align}
p(X = {x_i},\hat X = {x_j}) = p(X = {x_j},\hat X = {x_i}).
\end{align}

Since the achievability of the information rate-distortion-perception function ${R^{(I)}}(D,P)$ has not been proved yet, we analyze the relation between the rate-distortion functions $R(D,\infty )$ and $R(D,0)$. Assume a discrete memoryless stationary source $X$, let $Y = ({X_1},{X_2},...,{X_t})$ be a source sequence with length $t$. Consider a compression task $Y \to Z \to \hat Y$, where $Z$ is the output of the encoder and $\hat Y$ is the output of the decoder. As defined in \cite{ThomasCover2nd}, $R(D,\infty )$ is the infimum of rates $R$ for a given distortion $D$, such that $(R,D)$ is achievable. Hence, $R(D,\infty )$ can be expressed as
\begin{align}
R(D,\infty ) = \mathop {\inf }\limits_t {F_t}(D,\infty ),
\end{align}
where
\begin{align}
\begin{split}
&{F_t}(D,\infty ): = \mathop {\min }\limits_{{p_{Z|Y}},{p_{\hat Y|Z}}} \frac{{\rm{1}}}{t}H(Z)\\
&\text{s.t.}\quad \frac{{\rm{1}}}{t}\mathbb{E}[\Delta (Y,\hat Y)] \le D.\label{tRDfunction}
\end{split}
\end{align}
Considering the condition of perfect perception constraint, $R(D,0)$ can be similarly expressed as \cite{2018RDPtradeoff}
\begin{align}
R(D,0) = \mathop {\inf }\limits_t {F_t}(D,0),
\end{align}
where
\begin{align}
\begin{split}
&{F_t}(D,0): = \mathop {\min }\limits_{{p_{Z|Y}},{p_{\hat Y|Z}}} \frac{{\rm{1}}}{t}H(Z)\\
&\text{s.t.}\quad \frac{{\rm{1}}}{t}\mathbb{E}[\Delta (Y,\hat Y)] \le D,\ d({p_Y},{p_{\hat Y}}) \le 0.\label{tRDP0function}
\end{split}
\end{align}
For the MSE distortion measure, we have the following result (proof is given in the supplementary material).

\textbf{Theorem 1.} \textit{Let $X$ be a discrete memoryless stationary source and $Y = ({X_1},{X_2},...,{X_t})$ be a source sequence with length $t$. For a compression task $Y \to Z \to \hat Y$, where $Z$ and $\hat Y$ are the outputs of the encoder and decoder, respectively. Then, to any $t$, if $\Delta$ is the squared-error distortion, there exists an optimal encoder-decoder pair to \eqref{tRDP0function} satisfying}
\begin{align}
{p_{Y|Z}} = {p_{\hat Y|Z}}.\label{theorem1}
\end{align}
Theorem 1 indicates that when the distortion measure is MSE, replacing the perfect perception constraint $d({p_X},{p_{\hat X}}) \le 0$ in \eqref{tRDP0function} with the conditional distribution constraint \eqref{theorem1} does not change the optimal objective value. This is because under the condition that the data source is memoryless and stationary, it follows from theorem 1 that there exists an optimal solution to \eqref{tRDP0function} satisfying \eqref{theorem1}, under which the distributions of $X$ and $\hat X$ are the same. Therefore, analyzing the relation between ${F_t}(D,\infty )$ and ${F_t}(D,0)$ is equivalent to analyzing the relation between ${F_t}(D,\infty )$ and
\begin{align}
\begin{split}
&{F_t}(D,0) = \mathop {\min }\limits_{{p_{Z|Y}},{p_{\hat Y|Z}}} \frac{{\rm{1}}}{t}H(Z)\\
&\text{s.t.}\quad \frac{{\rm{1}}}{t}\mathbb{E}[\Delta (Y,\hat Y)] \le D,\ {p_{Y|Z}} = {p_{\hat Y|Z}}.\label{FtD02}
\end{split}
\end{align}
For any encoder-decoder pair satisfying \eqref{theorem1}, the MSE distortion between $Y$ and $\hat Y$ can be expressed as (proof is given in the supplementary material)
\begin{align}
\frac{{\rm{1}}}{t}\mathbb{E}\left[ {{{\left\| {Y - \hat Y} \right\|}^2}} \right] = \frac{{\rm{2}}}{t}\mathbb{E}\left[ {{{\left\| {Y - \mathbb{E}[Y|Z]} \right\|}^2}|Z} \right],
\end{align}
which substituted into \eqref{FtD02} yields
\begin{align}
\begin{split}
&{F_t}(D,0) = \mathop {\min }\limits_{{p_{Z|Y}},{p_{\hat Y|Z}}} \frac{{\rm{1}}}{t}H(Z)\\
&\text{s.t.}\quad \frac{{\rm{2}}}{t}\mathbb{E}\left[ {{{\left\| {Y - \mathbb{E}[Y|Z]} \right\|}^2}|Z} \right] \le D,\ {p_{Y|Z}} = {p_{\hat Y|Z}}.\label{FtD03}
\end{split}
\end{align}
It can be seen that the objective value of \eqref{FtD03} is independent on the decoder ${p_{\hat Y|Z}}$. Hence, with MSE distortion, the encoder of \eqref{FtD02} can be optimized separately as
\begin{align}
\begin{split}
&{F_t}(D,0) = \mathop {\min }\limits_{{p_{Z|Y}}} \frac{{\rm{1}}}{t}H(Z)\\
&\text{s.t.}\ \frac{{\rm{2}}}{t}\mathbb{E}\left[ {{{\left\| {Y - \mathbb{E}[Y|Z]} \right\|}^2}|Z} \right] \le D.\label{FtD04}
\end{split}
\end{align}
In \eqref{FtD04}, the constraint ${p_{Y|Z}} = {p_{\hat Y|Z}}$ is removed since it does not affect the objective under the constraint $\frac{{\rm{2}}}{t}\mathbb{E}\left[ {{{\left\| {Y - \mathbb{E}[Y|Z]} \right\|}^2}|Z} \right] \le D$.

Next, we consider the lossy compression task without perception constraint. Given a compressed representation $z$ and when the distortion measure is MSE, the output of an optimal decoder is $\mathbb{E}[Y|z]$. Accordingly, the expected distortion can be expressed as
\begin{align}
\mathbb{E}[\Delta (Y,\mathbb{E}[Y|Z])] = \frac{1}{t}\mathbb{E}\left[ {{{\left\| {Y - \mathbb{E}[Y|Z]} \right\|}^2}|Z} \right],
\end{align}
with which formulation \eqref{tRDfunction} can be rewritten as
\begin{align}
\begin{split}
&{F_t}(D,\infty) = \mathop {\min }\limits_{{p_{Z|Y}}} \frac{{\rm{1}}}{t}H(Z)\\
&\text{s.t.}\quad \frac{{\rm{1}}}{t}\mathbb{E}\left[ {{{\left\| {Y - \mathbb{E}[Y|Z]} \right\|}^2}|Z} \right] \le D.\label{FtD2}
\end{split}
\end{align}
Then, it follows from \eqref{FtD04} and \eqref{FtD2} that ${F_t}(D,0) = {F_t}(D/2,\infty )$ for any $t$, which implies
\begin{align}
\mathop {\inf }\limits_t {F_t}(D,0) = \mathop {\inf }\limits_t {F_t}(D/2,\infty ).
\end{align}
Thus, we have the following result.

\textbf{Theorem 2.} \textit{Suppose that $\Delta$ is the squared-error distortion, then $R(D,P)$ satisfies}
\begin{align}
R(D,0) = R\left( {\frac{1}{2}D, + \infty } \right).\label{theorem2}
\end{align}
Theorem 2 indicates that, when the distortion is measured by MSE, and for fixed bit rate, the lowest achievable distortion under perfect perception constraint is 2-fold that under no perception constraint. In other words, the cost of attaining perfect perceptual quality is exactly a doubling of the lowest achievable MSE distortion.

Moreover, \eqref{FtD04} and \eqref{FtD2} also imply that, with distortionconstrained by $D/2$, an optimal encoder under no perception constraint is also an optimal encoder under perfect perception constraint with distortion constrained by $D$.

\textbf{Theorem 3.} \textit{When the distortion is measured by MSE, an optimal encoder to \eqref{tRDfunction} (without perception constraint) with distortion constrained by $D/2$ is also an optimal encoder to \eqref{tRDP0function} (with perfect perception constraint) with distortion constrained by $D$.}

Theorem 3 implies that the rate-distortion function with perfect perception constraint is achievable. Besides, it sheds some light on how to optimize an encoder-decoder pair under perfect perception constraint. Specifically, the encoder can be independently optimized without considering perception constraint. In light of this, we can first optimize an encoder with only MSE constraint, and then fix the encoder and optimize the decoder to satisfy \eqref{theorem1}, which would double the MSE distortion and has been proven to be an optimal decoder that can achieve the bound in Theorem 2. Such a framework is detailed in the next subsection.

\begin{figure}[!t]
	\vskip 0.2in
	\begin{center}
		\centerline{\includegraphics[width=0.8\columnwidth]{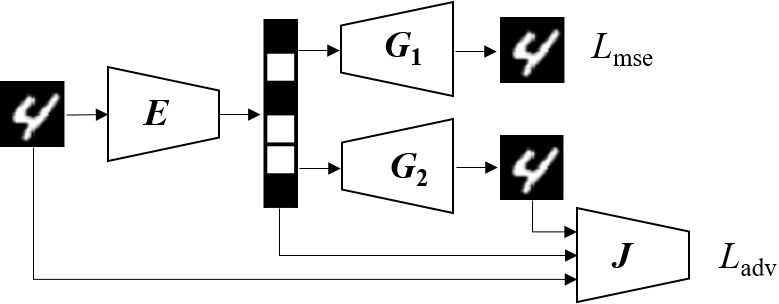}}
		\caption{\textbf{The proposed trianing framework.} First, the encoder-decoder pair $(E,G_1)$ is trained by MSE loss. Then $G_2$ is trained by adversarial loss conditioned on $E$.}
		\label{TrainingFramework}
	\end{center}
	\vskip -0.2in
\end{figure}

\subsection{A Framework for Perfect Perceptual Reconstruction in Lossy Compression}

Based on the above results, we further propose a framework for training an encoder-decoder pair to achieve perfect perceptual reconstruction in lossy compression. The overall architecture is shown in Figure~\ref{TrainingFramework}, which includes an encoder $E$, two decoders $G_1$, $G_2$, and a discriminator $J$. The desired encoder-decoder pair is $(E,G_2)$, which is obtained by a training procedure with two steps:

\setlength{\hangindent}{1em}
\textit{i) Encoder optimization:} $(E,G_1)$ are optimized only in terms of MSE distortion, e.g. by minimizing MSE loss without considering the perception constraint. From Theorem 3, such an optimized encoder $E$ is also an optimal encoder under perfect perception constraint.

\setlength{\hangindent}{1em}
\textit{ii) Decoder optimization:} Fixing the optimized encoder $E$ in the first step, the decoder $G_2$ is optimized via adversarial training, e.g. iteratively optimize $G_2$ and the discriminator $J$. In this step, the goal is to minimize the divergence between ${p_{Y|Z}}$ and ${p_{{\hat Y}|Z}}$. To achieve this, we use a conditional discriminator \cite{2014ConGAN} which takes both the data $(Y$ or $\hat Y)$ and the bit stream $Z$ as the input.

Our proposed framework is different from traditional DAL framework as follows. As discussed in Section 2.3, DAL involves a balance between the distortion and adversarial losses, and is difficult to achieve the lowest distortion under perfect perception constraint. Unlike the discriminator in DAL method being used to discriminate whether its input is real, the discriminator in our framework not only discriminate whether input data is real but also discriminate whether the input data is consistent with the bit stream $Z$. By this, though without distortion constraint, $G_2$ is trained to decode data consistent with the input rather than randomly generate a realistic output. This results in an advantage that the proposed framework avoids the balance between the distortion and adversarial losses in DAL, and can achieve the lowest MSE distortion under perfect perception constraint.

To optimize the decoder to satisfy \eqref{theorem1}, we add a random noise input in the decoder $G_2$. This is a commonly used trick to achieve high perceptual quality \cite{2019Rethinking, 2019Extreme}, In our framework, \eqref{theorem1} holds when $(G_2,J)$ are optimal, which means the output of the decoder would have perfect perceptual quality.

\section{Experimental Illustration}

\begin{figure}[!t]
	\vskip 0.2in
	\begin{center}
		\centerline{\includegraphics[width=\columnwidth]{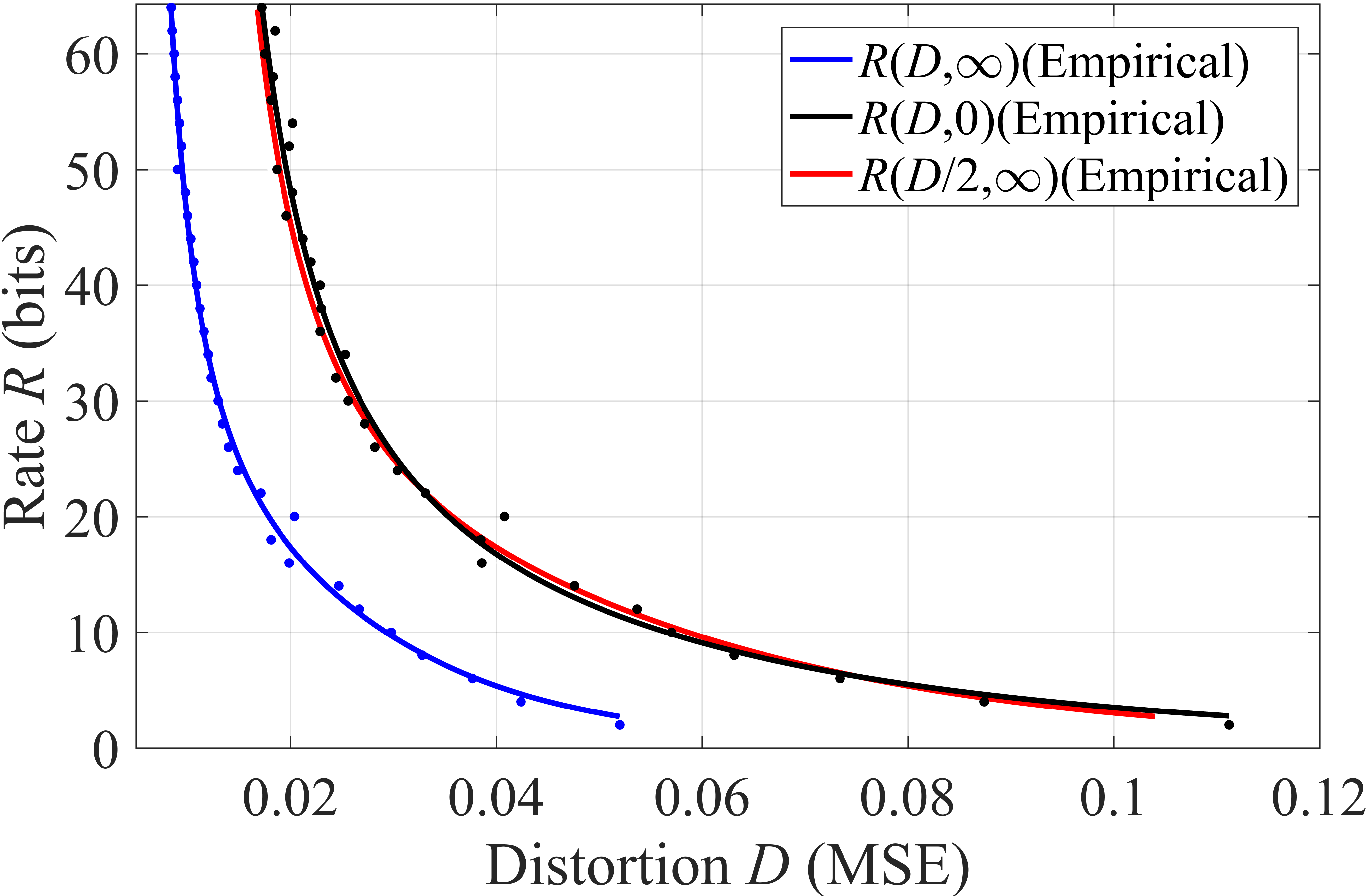}}
		\caption{Empirically fitted rate-distortion functions in the cases with or without perception constraint.}
		\label{EmpiricalRDP}
	\end{center}
	\vskip -0.2in
\end{figure}

We demonstrate the theoretical finding and the effectiveness of the proposed framework on the MNIST dataset \cite{1998Gradient}. Note that the rate-distortion-perception function does not have an analytical expression for a general data distribution, here we empirically demonstrate the 2-fold relation in Theorem 2 by experiment. As the compression task on the MNIST data is relatively simple, and the proposed framework can theoretically achieve the lowest MSE distortion with perfect perceptual quality as discussed in Section 3.2, an implemantation using DNN can be expected to closely approach the rate-distortion curve under perception constraint. In this context, we can compare the empirical rate-distortion curve of the proposed framework with that under no perception constraint to demonstrate the derived 2-fold relation. Moreover, we demonstrate the superiority of the proposed framework by comparing it with the DAL method. The code is available online\footnote{https://github.com/ZeyuYan/Perceptual-Lossy-Compression}.

In the proposed framework, $E$, $G_1$, $G_2$ and $J$ are convolutional neural networks (CNN). The encoder $E$ maps a $32 \times 32$ image into a $d \times 1$ vector with each element be a quantized integer. To preserve gradient for backpropagation, the elements are quantized by
\begin{align}
\hat z = z{\rm{  +  sg[}}Q(z) - z{\rm{]}},
\end{align}
where $Q( \cdot )$ is a quantization operator and ${\rm{sg[}} \cdot {\rm{]}}$ is the stopgradient operator \cite{2017Neural, 2019VQVAE2}. The bit rate is controlled by limiting the dimension and quantization level of encoder output. The output is quantized into binary values and correspondingly the bit rate is bounded by $d$.

Our model is trained by two steps as presented in Section 3.2, in which $(E,G_1)$ is firstly optimized as
\begin{align}
\mathop {{\rm{min}}}\limits_{E,{G_1}} \mathbb{E} {\left\| {X - {G_1}(E(X))} \right\|^2}.
\end{align}
Then, with $E$ fixed, $(G_2,J)$ is optimized as
\begin{align}
\mathop {\min }\limits_{J\in \mathcal{F}} \mathop {\max }\limits_{{G_2}} \mathbb{E}[J({G_2}(E(X)),E(X))] - \mathbb{E}[J(X,E(X))],\notag
\end{align}
where $\mathcal{F}$ denotes the bounded 1-Lipschitz functions. $G_2$ is trained to generate images of which the distribution is the same as that of $X$. The input of $G_2$ is a $100 \times 1$ vector, in which the first $d\ (d<100)$ elements are the output of $E$ and the other elements are Gaussian noise. When $d=0$, $(G_2,J)$ degenerates to a pure generative adversarial network. Extra experiments show that, with different dimensions of the noise vector, the MSE curves converge to almost the same value. To stabilize the process of adversarial training, WGAN-gp \cite{2017WGAN, 2017Wasserstein} is employed, in which a gradient penalty is added into the loss function of $J$ as
\begin{align}
\mathbb{E}[J({G_2}(E(X)),E(X))] - \mathbb{E}[J(X,E(X))] + {\lambda _{gp}}{L_{gp}}, \notag
\end{align}
where $L_{gp}$ is the gradient penalty term and $\lambda _{gp}$ is a parameter set to 10. We use a pre-training scheme to address the degeneration problem in adversarial training, which is detailed in the supplementary material.

\begin{figure*}[!t]
	\vskip 0.2in
	\begin{center}
		\subfigure[2 bits]{
	        \begin{minipage}[t]{0.5\columnwidth}
		        \centering
	            \includegraphics[width=\columnwidth]{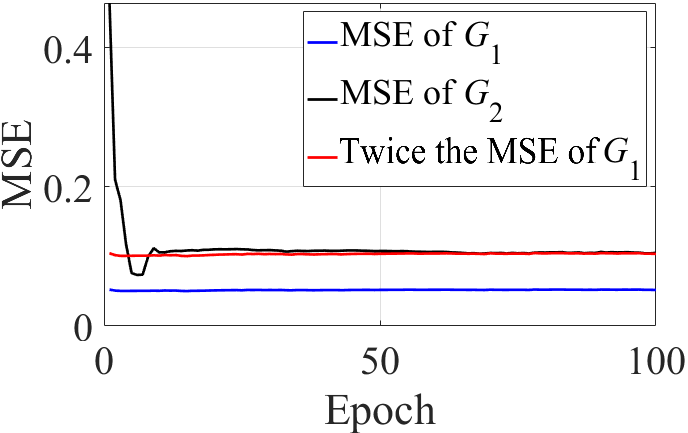}
	        \end{minipage}%
        }%
		\subfigure[4 bits]{
			\begin{minipage}[t]{0.5\columnwidth}
				\centering
				\includegraphics[width=\columnwidth]{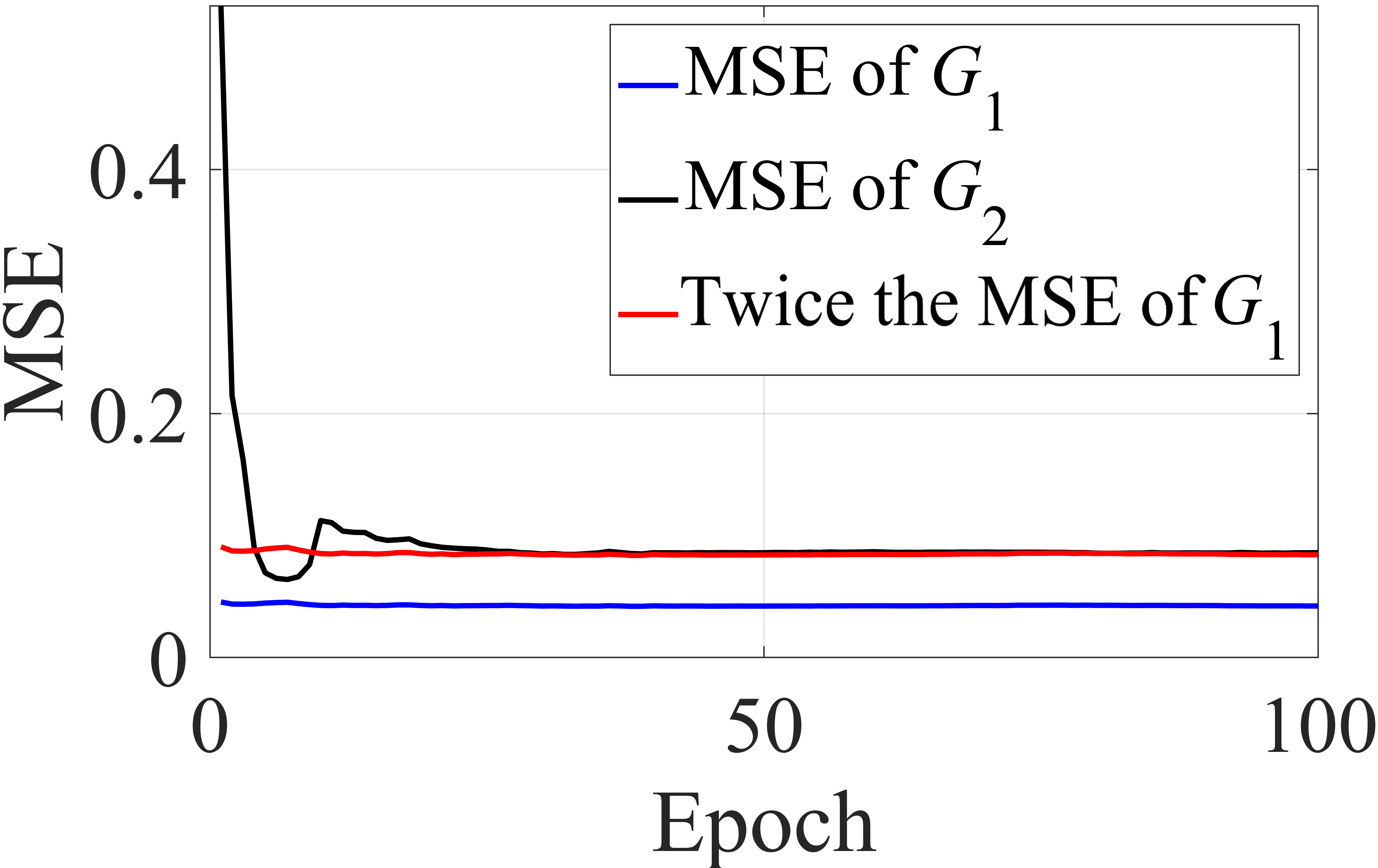}
			\end{minipage}%
		}%
		\subfigure[8 bits]{
			\begin{minipage}[t]{0.5\columnwidth}
				\centering
				\includegraphics[width=\columnwidth]{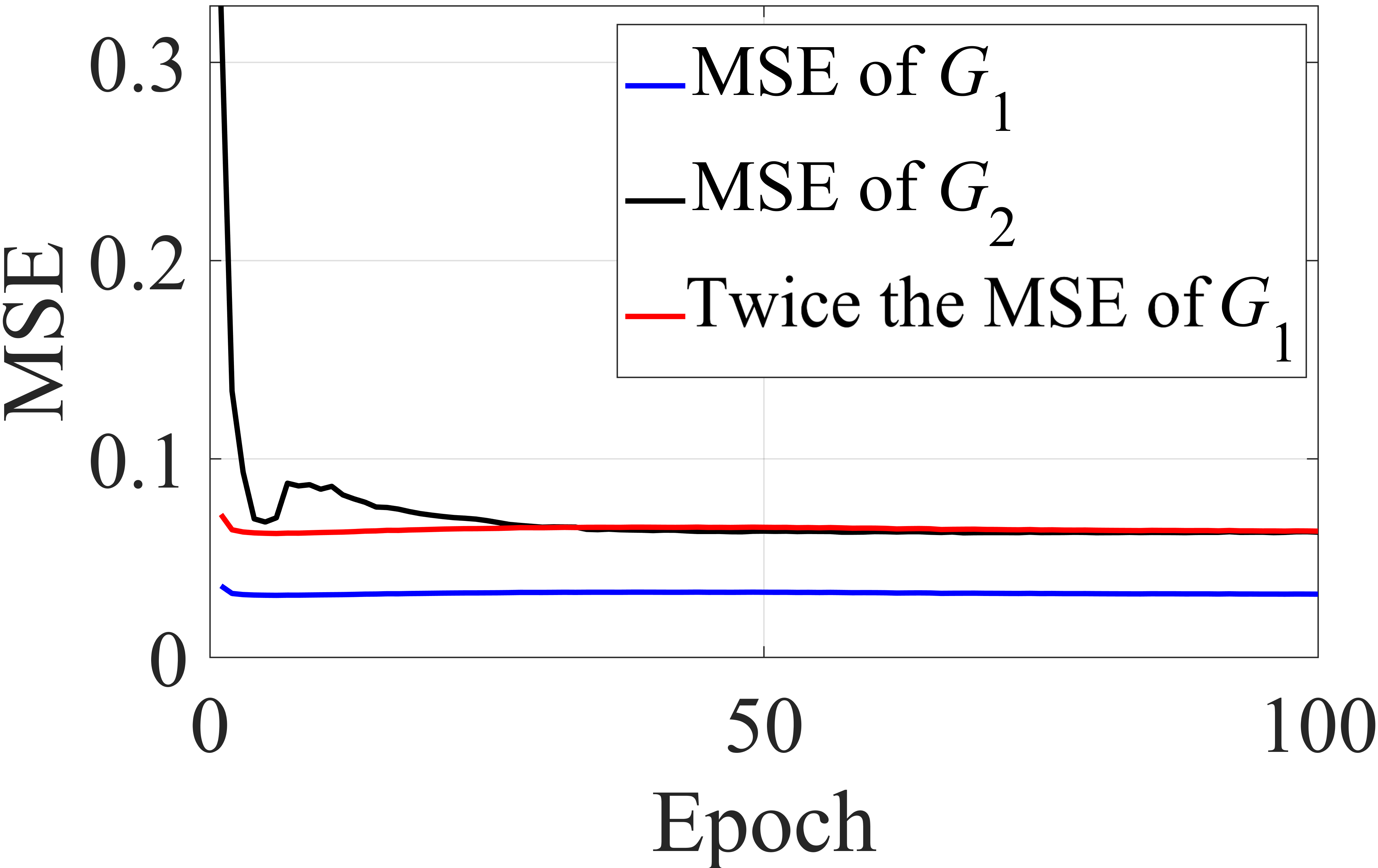}
			\end{minipage}
		}
	
		\subfigure[16 bits]{
			\begin{minipage}[t]{0.5\columnwidth}
				\centering
				\includegraphics[width=\columnwidth]{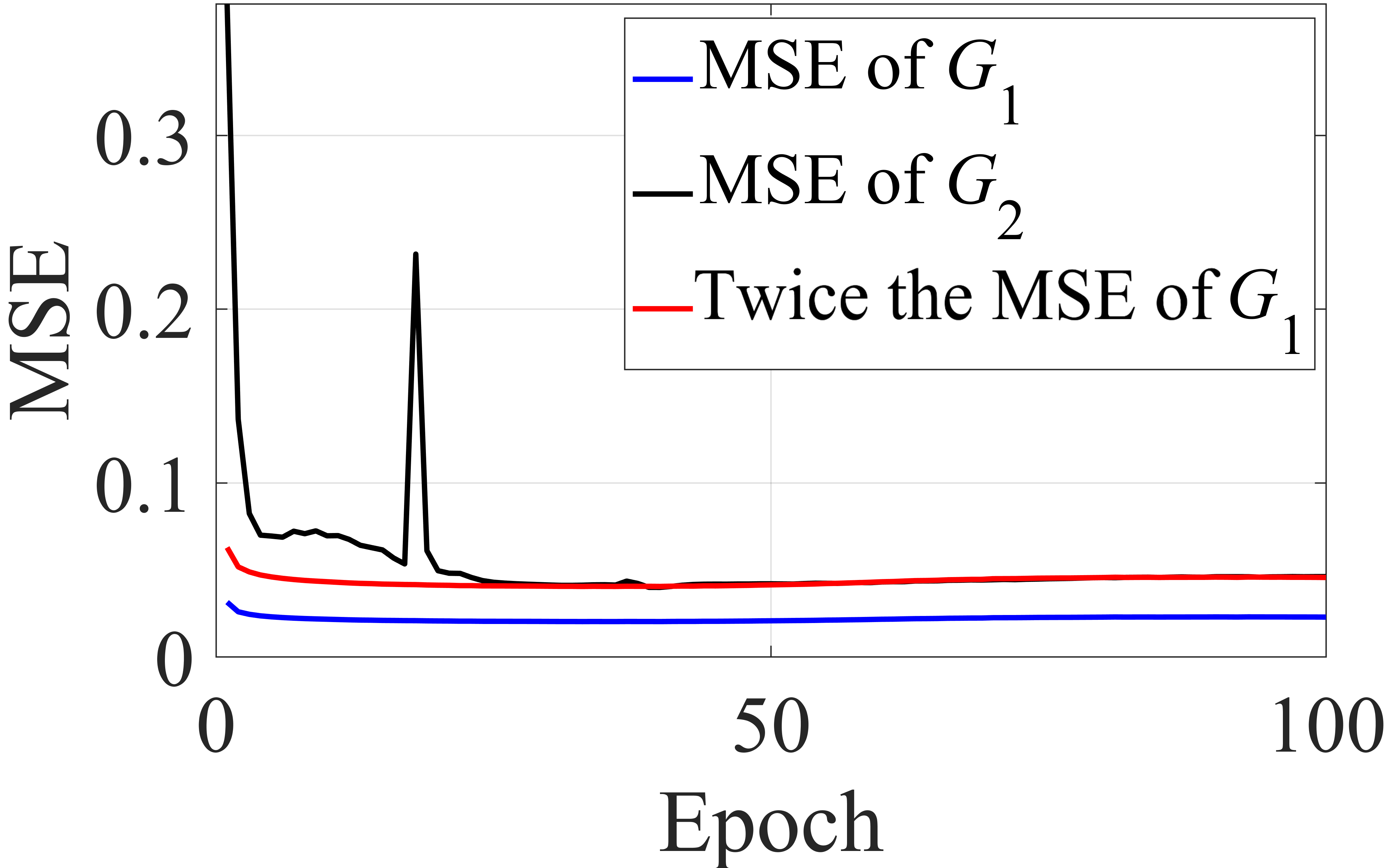}
			\end{minipage}%
		}%
		\subfigure[32 bits]{
			\begin{minipage}[t]{0.5\columnwidth}
				\centering
				\includegraphics[width=\columnwidth]{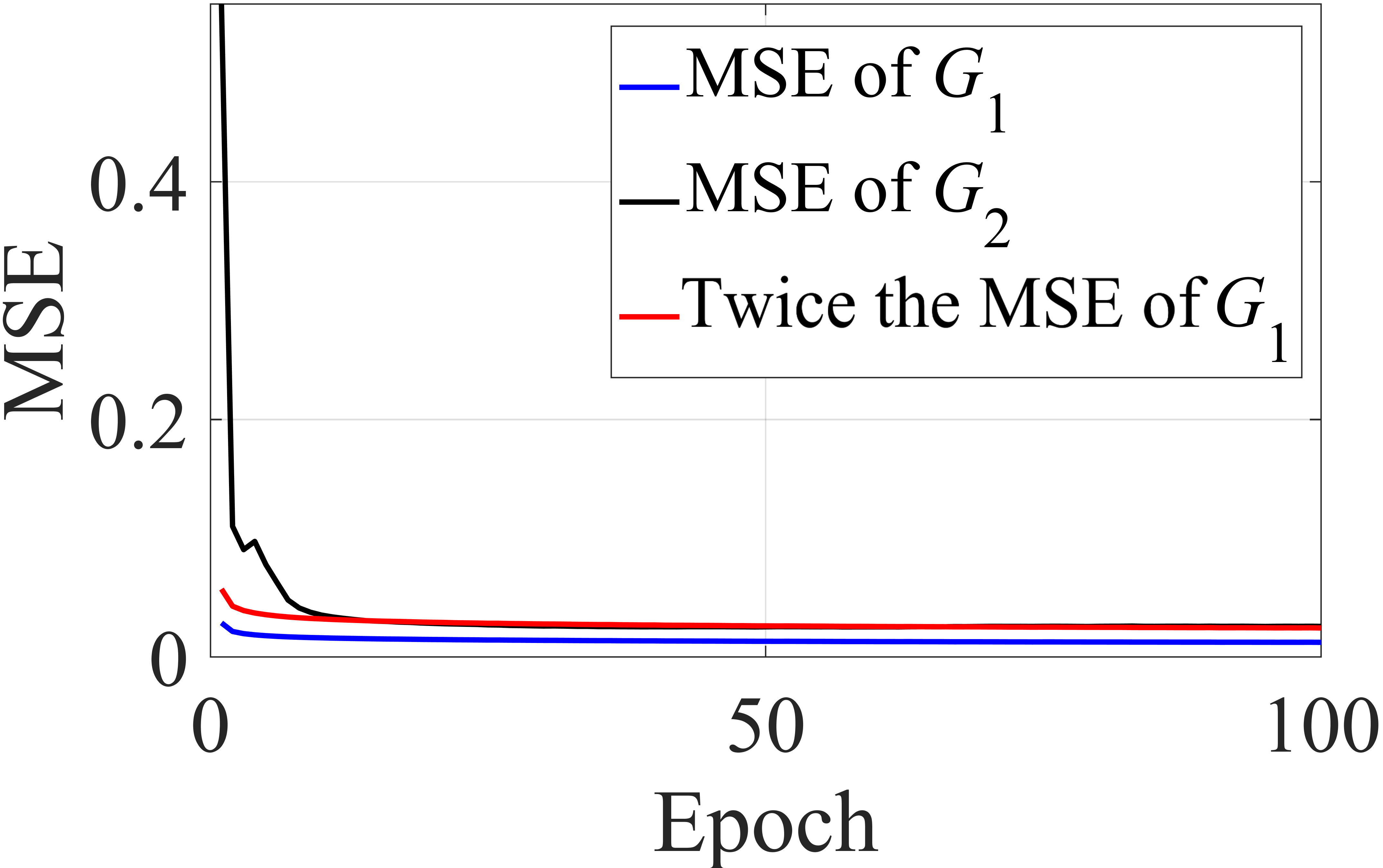}
			\end{minipage}%
		}%
		\subfigure[64 bits]{
			\begin{minipage}[t]{0.5\columnwidth}
				\centering
				\includegraphics[width=1\columnwidth]{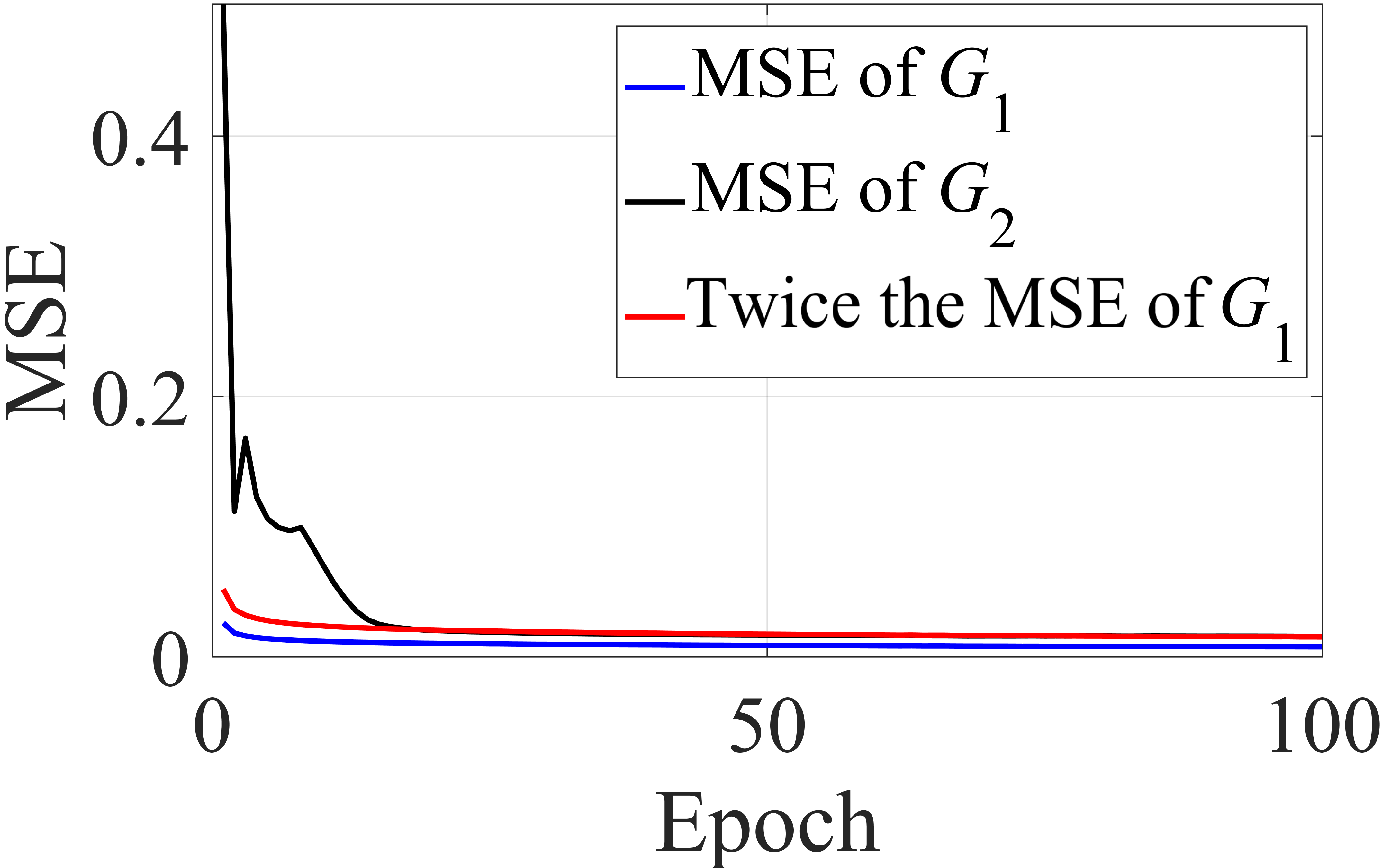}
			\end{minipage}
		}
		
		\caption{MSE loss versus training epoch for different bit-rates.}
		\label{Trainloss}
	\end{center}
	\vskip -0.2in
\end{figure*}

\subsection{Rate-Distortion Tradeoff with or without Perception Constraint}

For 32 different bit rates, $R \in \{ 2,4,6, \cdots ,64\}$, we train 32 encoder-decoder pairs $(E,G_1)$ by minimizing MSE-only loss and another 32 pairs $(E,G_2)$ by our proposed framework. Figure~\ref{EmpiricalRDP} shows the fitted empirical rate-distortion curves of the encoder-decoder pairs. In Figure~\ref{EmpiricalRDP}, $R(D,\infty)$ and $R(D,0)$ are the fitting results, whilst $R(D/2,\infty)$ is a scaled version of the fitted $R(D,\infty)$ curve. Clearly, the empirical rate-distortion curves are monotonically non-increasing and convex, which is consistent with the theoretical properties of $R(D,P)$ \cite{2019Rethinking}. Interestingly, $R(D/2,\infty)$ closely approaches $R(D,0)$, which empirically demonstrates the result given by Theorem 2, i.e., the 2-fold relation between the lowest achievable MSE distortion under perfect perception constraint and that under no perception constraint.

Figure~\ref{Trainloss} presents the MSE loss in training $G_1$ and $G_2$ versus training epoch for six rate cases $R \in \{ 2,4,8,16,32,64\}$. For each case, the curve which is twice the amplitude of the MSE of $G_1$ is also plotted for reference. It can be seen that, without perception constraint, the MSE loss of $G_1$ converges rapidly in all bit-rate cases. In comparison, the MSE loss of $G_2$ converges slower, which is due to the fact that $G_2$ is jointly trained with $J$ in an alternating manner. Although $G_2$ is trained only using an adversarial loss and without using MSE loss, it closely converges to the 2-fold MSE curve of $G_1$ in each case, which accords well with Theorem 2.

\begin{figure}[!t]
	\vskip 0.2in
	\begin{center}
		\centerline{\includegraphics[width=0.8\columnwidth]{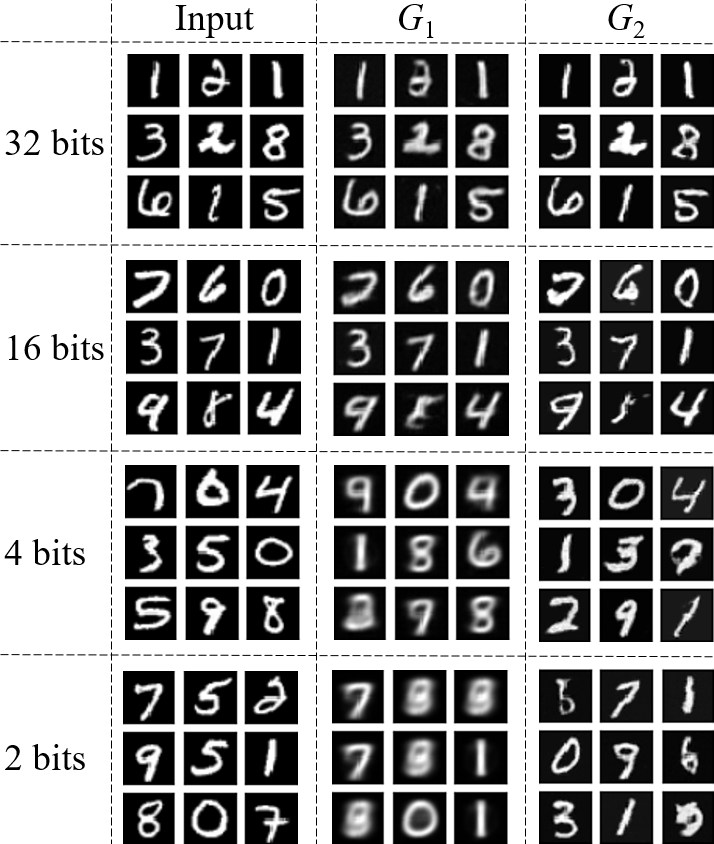}}
		\caption{Visual comparison between the two cases with or without perception constraint.}
		\label{Outputimage}
	\end{center}
	\vskip -0.2in
\end{figure}

\begin{figure*}[!t]
\vskip 0.2in
\begin{center}
	\centerline{\includegraphics[width=1.58\columnwidth]{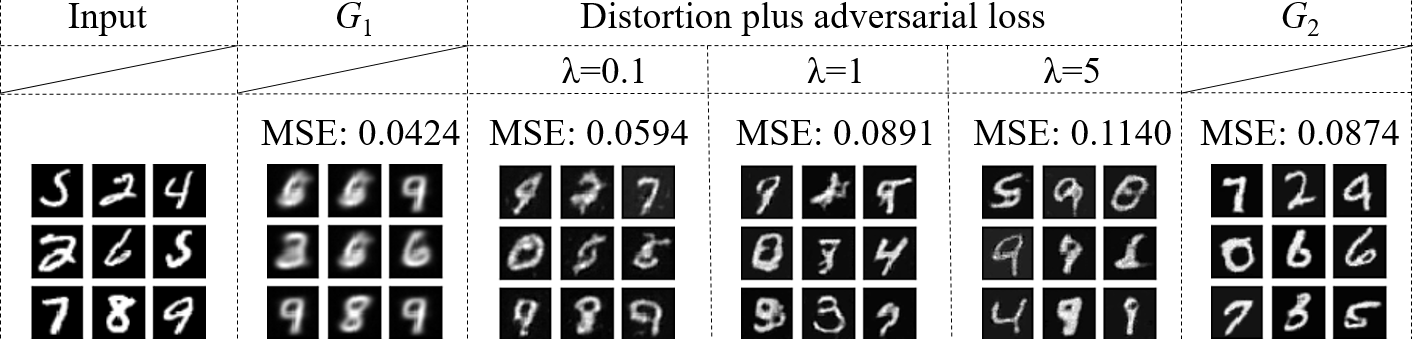}}
	\caption{Performance comparison between the traditional distortion plus adversarial loss based method and the proposed method.}
	\label{ComparewithDAL}
\end{center}
\vskip -0.2in
\end{figure*}

Figure~\ref{Outputimage} shows some samples of the reconstructed images by $G_1$ and $G_2$ under different bit rates. Obviously, the output of $G_2$ is clearer and sharper than that of $G_1$. As the decrease of bit rate, the output of $G_1$ becomes more blurry and unrecognizable, especially when bit rate is less than 4, which is due to the information lost in the encoding step. In comparison, the output of $G_2$ does not suffer from such deteriorating problem, which is clear and recognizable even when the bit rate is 2. However, for fixed bit rate, better perceptual quality would lead to larger distortion. Hence clear output images do not necessarily mean correct numbers. In principle, $G_2$ additionally takes a noise input to generate details, which would change the typeface, inclination, thickness of the original input and may even reconstruct a completely different but clear number.

Next, we compare our method with the DAL method, of which the parameter $\lambda$ balances between the MSE distortion loss and the adversarial loss. Based on DAL for different values of $\lambda$, we train a number of encoder-decoder pairs for bit rate $R=4$. Figure~\ref{ComparewithDAL} compares the results of $G_1$, $G_2$ and the DAL method with selected $\lambda  \in \{ 0.1,1,5\} $, including the MSE and output image samples. It can be seen that, for the DAL method, as the increase of $\lambda$, the perceptual quality improves but in the meantime the distortion increases. Our method has better peceptual quality than the DAL method even when the MSE of DAL is larger than that of $G_2$. As discussed in Section 2.3 and 3.2, while our method can achieve the lowest distortion under perfect perception constraint, the DAL method cannot.

\subsection{Related Works}

Our work is closely related to \cite{2019Rethinking} but different from it in the following aspects. First, the analysis in \cite{2019Rethinking} is based on constructing an encoder-decoder pair that can attain perfect perception quality, e.g., by concatenating a post-processing perceptual mapping after an optimal encoder-decoder pair under MSE measure. Since the constructed compression system can attain perfect perception quality with a doubling of the MSE distortion, in theory the lowest achievable MSE distortion under perfect perception constraint should be no more than twice that under no perception constraint. In contrast, we derive the rate-distortion bound under perfect perception constraint through a completely different line of analysis, via analyzing the lossy compression formulation itself as presented in Section 3.1.  Second, our result deterministically shows that the lowest achievable distortion under perfect perception constraint is exactly 2-fold that under no perception constraint. Third, we show that, to achieve the lowest MSE distortion under perfect perception constraint, perceptual loss is unnecessary for training the encoder whilst distortion loss, either on pixel or deep features, is unnecessary for training the decoder. Fourth, we propose a training framework that can achieve the lowest MSE distortion under perfect perception constraint at a given bit rate.

Like \cite{2018Unreasonable, 2017PhotoRealistic}, our method is also based on GAN. However, unlike the formers using a distortion plus adversarial loss to train the encoder and decoder in an end-to-end manner, our method first uses a MSE loss to train the encoder and, then, uses an adversarial loss to train the decoder conditioned on the so obtained encoder. The superiority of the new method is discussed in Section 2.3 and 3.2. Recently, a conditioned discriminator similar to that in our framework has been employed in \cite{2020HighFidelity} to achieve high-fidelity image compression. Our work provides a theoretical foundation of the conditioned discriminator. The work \cite{2020HighFidelity} still uses a distortion plus adversarial loss, while we show that the decoder can be
trained to achieve perfect perceptual reconstruction using
adversarial-only loss and adversarial loss is unnecessary when training encoder.

We leave the application of the proposed framework to color image compression to future work, as it has a high requirement on computation hardware and poses a challenge in tuning a big GAN model.

\section{Conclusion}


We analyzed the effect of perception constraint on the rate-distortion function in lossy compression. We proved that, for fixed bit rate, the cost of imposing a perfect perception constraint is exactly a doubling of the lowest achievable MSE. The analysis also provided new insights on how to build a training framework for perfect perception reconstruction in lossy compression. Accordingly, we proposed a framework for training an encoder-decoder pair to achieve the lowest MSE under perfect perception constraint. Experimental results well verified the theoretical finding and demonstrated the superiority of the new framework over the traditional distortion plus adversarial loss based framework.

\bibliography{mypaper}
\bibliographystyle{icml2021}

\appendix
\onecolumn
\icmltitle{On Perceptual Lossy Compression: The Cost of Perceptual\\
	Reconstruction and An Optimal Training Framework\\
	Supplementary Material}







This supplemental material first provides the proof of lemma and theorem 1, and then gives the derivation of equation (20). Finally, degeneration problem is discussed and we provide a pre-training trick to solve it.

\section{Proof of Lemma 1}
Suppose ${{\bf{B}}^*}$ is an optimal solution to (10), which satisfies $\left\langle {{\bf{W}},{{\bf{B}}^*}} \right\rangle  \le D$ and $\sum\limits_{i = 1}^m {b_{ij}^*}  = \sum\limits_{i = 1}^m {b_{ji}^*}  = p(X = {x_j}), {\rm{1}} \le j \le m$. Since $\Delta $ is symmetric, ${w_{ij}} = \Delta ({x_i},{x_j}) = \Delta ({x_j},{x_i}) = {w_{ji}}$ holds for any ${\rm{1}} \le i,j \le m$, which means ${\bf{W}} = {{\bf{W}}^T}$. Thus, we have
\begin{align}
\left\langle {{\bf{W}},{{\bf{B}}^{*T}}} \right\rangle  = \left\langle {{{\bf{W}}^T},{{\bf{B}}^{*T}}} \right\rangle  = \left\langle {{\bf{W}},{{\bf{B}}^*}} \right\rangle  \le D.
\end{align}
Denote the $(i,j)$-th element of ${{\bf{B}}^{*T}}$ by $b_{ij}'$,  it follows that $b_{ij}' = b_{ji}^*$, so that
\begin{align}
\begin{split}
\sum\limits_{i = 1}^m {b_{ij}'}  = \sum\limits_{i = 1}^m {b_{ji}^*}  = p(X = {x_j})\\
\sum\limits_{i = 1}^m {b_{ji}'}  = \sum\limits_{i = 1}^m {b_{ij}^*}  = p(X = {x_j}).
\end{split}
\end{align}
Then, it is easy to see that ${{\bf{B}}^{*T}}$ is also a feasible solution to (10).  Meanwhile, it can be justified that ${{\bf{B}}^{*T}}$ is also an optimal solution to (10) since the objective satisfies
\begin{align}
\begin{split}
{G_{{p_X}}}({{\bf{B}}^{*T}}) &= 2H(X) + \sum\limits_{i = 1}^m {\sum\limits_{j = 1}^m {b_{ij}'} \log b_{ij}'} \\
&= 2H(X) + \sum\limits_{i = 1}^m {\sum\limits_{j = 1}^m {b_{ji}^*} \log b_{ji}^*} \\
&= {G_{{p_X}}}({{\bf{B}}^*}).
\end{split}
\end{align}
Next, denote ${{\bf{B}}_0}: = ({{\bf{B}}^*}{\rm{ + }}{{\bf{B}}^{*T}}{\rm{)/2}}$, we show that ${G_{{p_X}}}({{\bf{B}}_0}) = {G_{{p_X}}}({{\bf{B}}^*})$. First, ${{\bf{B}}_0}$ is a feasible solution of (10) as it satisfies  the constraints
\begin{align}
\begin{split}
\left\langle {{\bf{W}},{{\bf{B}}_0}} \right\rangle  = \left\langle {{{\bf{W}}^T},\frac{{{{\bf{B}}^*}{\rm{ + }}{{\bf{B}}^{*T}}}}{2}} \right\rangle  = \frac{{\left\langle {{\bf{W}},{{\bf{B}}^*}} \right\rangle  + \left\langle {{\bf{W}},{{\bf{B}}^{*T}}} \right\rangle }}{2} \le D\\
\sum\limits_{i = 1}^m {{b_{0ij}}}  = \sum\limits_{i = 1}^m {\frac{{b_{ij}^* + b_{ij}'}}{2}}  = \frac{1}{2}\left( {\sum\limits_{i = 1}^m {b_{ij}^*}  + \sum\limits_{i = 1}^m {b_{ij}'} } \right) = p(X = {x_j})\\
\sum\limits_{i = 1}^m {{b_{0ji}}}  = \sum\limits_{i = 1}^m {\frac{{b_{ji}^* + b_{ji}'}}{2}}  = \frac{1}{2}\left( {\sum\limits_{i = 1}^m {b_{ji}^*}  + \sum\limits_{i = 1}^m {b_{ji}'} } \right) = p(X = {x_j}).
\end{split}
\end{align}
Meanwhile, the objective function ${G_{{p_X}}}({{\bf{B}}_0})$ can be expressed as
\begin{align}
\begin{split}
{G_{{p_X}}}({{\bf{B}}_0}) &= 2H(X) + \sum\limits_{i = 1}^m {\sum\limits_{j = 1}^m {{b_{0ij}}\log {b_{0ij}}} } \\
&= 2H(X) + \sum\limits_{i = 1}^m {\sum\limits_{j = 1}^m {\frac{{b_{ij}^* + b_{ij}'}}{2}\log \frac{{b_{ij}^* + b_{ij}'}}{2}} }.
\end{split}
\end{align}
Notice that the function $f(x) = x\log x$ is strictly convex in $(0,1)$. Thus we have
\begin{align}
\frac{{b_{ij}^* + b_{ij}'}}{2}\log \frac{{b_{ij}^* + b_{ij}'}}{2} \le \frac{1}{2}\left( {b_{ij}^*\log b_{ij}^* + b_{ij}'\log b_{ij}'} \right),
\end{align}
where the equality holds if and only if $b_{ij}^* = b_{ij}'$. Then, it follows that
\begin{align}
\begin{split}
{G_{{p_X}}}({{\bf{B}}_0}) &= 2H(X) + \sum\limits_{i = 1}^m {\sum\limits_{j = 1}^m {\frac{{b_{ij}^* + b_{ij}'}}{2}\log \frac{{b_{ij}^* + b_{ij}'}}{2}} }\\
&\le \frac{1}{2}\left[ {\left( {2H(X) + \sum\limits_{i = 1}^m {\sum\limits_{j = 1}^m {b_{ij}^*\log b_{ij}^*} } } \right) + \left( {2H(X) + \sum\limits_{i = 1}^m {\sum\limits_{j = 1}^m {b_{ij}'\log b_{ij}'} } } \right)} \right]\\
&= \frac{1}{2}\left[ {{G_{{p_X}}}({{\bf{B}}^*}) + {G_{{p_X}}}({{\bf{B}}^{*T}})} \right] = {G_{{p_X}}}({{\bf{B}}^*}). \label{OFvalueofB0}
\end{split}
\end{align}
Recall that ${{\bf{B}}^*}$ is an optimal solution, hence ${G_{{p_X}}}({{\bf{B}}^*}) \le {G_{{p_X}}}({{\bf{B}}_0})$,  which together with \eqref{OFvalueofB0} leads to ${G_{{p_X}}}({{\bf{B}}_0}) = {G_{{p_X}}}({{\bf{B}}^*})$. Thus, ${{\bf{B}}_0}$ is an optimal solution and for any ${\rm{1}} \le i,j \le m$ we have
\begin{align}
\frac{{b_{ij}^* + b_{ij}'}}{2}\log \frac{{b_{ij}^* + b_{ij}'}}{2} = \frac{1}{2}\left( {b_{ij}^*\log b_{ij}^* + b_{ij}'\log b_{ij}'} \right),
\end{align}
Furthermore, since $f(x) = x\log x$ is strictly convex in $(0,1)$, we have $b_{ij}' = b_{ij}^*$ for any ${\rm{1}} \le i,j \le m$ and hence ${{\bf{B}}^*} = {{\bf{B}}^{*T}}$, which finally results in Lemma 1.

\section{Proof of Theorem 1}

Let $X$ be a memoryless stationary source, $Y = ({X_1},{X_2},...,{X_t})$ be a source sequence of length $t$, $L$ and $Q$ be the encoder and decoder, respectively, with which the compressed representation is $Z=L(Y)$ and the output of the encoder is ${\hat Y}=Q(Z)$. Since  $F_t(D,0)$ defined in (15) is non-increasing on $D$, in the case of squared-error distortion, we consider its inverse form for convenience as
\begin{align}
\begin{split}
\mathop {\min }\limits_{L,Q} \frac{1}{t}&\mathbb{E}\left[ {{{\left\| {Y - \hat Y} \right\|}^2}} \right]\\
s.t.\quad&{\rm{ }}Z = L(Y),{\rm{ }}\hat Y = Q(Z),\\
&H{\rm{(}}Z{\rm{)}} \le tR{\rm{, }}d({p_Y},{p_{\hat Y}}) \le 0, \label{B1}
\end{split}
\end{align}
which minimizes the MSE distortion under constraints on the average bit-rate and distribution divergence (perception quality).

For convenience in the sequel analysis, we define the joint distribution matrix of $Y$ and $Z$ as ${\bf{L}} \in {\mathbb{R}^{m \times n}}$ with the $(i,j)$-th element being ${l_{i,j}} = {p_{Y,Z}}({y_i},z_j^{})$. Similarly, we define the joint distribution matrix of $\hat Y$ and $Z$ as ${\bf{Q}} \in {\mathbb{R}^{m \times n}}$ with the $(i,j)$-th element being ${q_{i,j}} = {p_{\hat Y,Z}}({y_i},z_j^{})$. In fact, $\bf{L}$ and $\bf{Q}$ are the joint distribution matrices of the encoder and decoder, respectively.

Next we show that for any optimal encoder-decoder pair $({L^{\rm{*}}},{Q^{\rm{*}}})$ to \eqref{B1} with joint distribution matrices $({{\bf{L}}^{\rm{*}}},{{\bf{Q}}^{\rm{*}}})$, the  encoder-decoder pairs with joint distribution matrices $({{\bf{L}}^{\rm{*}}},{{\bf{L}}^{\rm{*}}})$ and $({{\bf{Q}}^{\rm{*}}},{{\bf{Q}}^{\rm{*}}})$ are also optimal to\eqref{B1}.

First, for an optimal encoder-decoder pair $({L^{\rm{*}}},{Q^{\rm{*}}})$ to \eqref{B1}, we have $H({L^{\rm{*}}}(Y)) \le tR$. Let the alphabet of $Y$ be ${\rm{\{ }}{y_1}{\rm{,}}{y_2},...,{y_m}{\rm{\} }}$ and the alphabet of $Z$ be ${\rm{\{ }}z_1{\rm{,}}z_2,...,z_n{\rm{\} }}$, and suppose that $p({L^{\rm{*}}}(Y) = z_j) = {h_j}, 1 \le j \le n$. Then we consider the following formulation
\begin{align}
\begin{split}
\mathop {\min }\limits_{L,Q} \frac{{\rm{1}}}{t}&\mathbb{E}{\rm{[}}{\left\| {Y - \hat Y} \right\|^2}{\rm{]}}\\
s.t.\quad &Z = L(Y),\ \hat Y = Q(Z),\ d({p_Y},{p_{\hat Y}}) \le 0,\\
&{p_Z}(z_j^{}) = {h_j},\ 1 \le j \le n. \label{B2}
\end{split}
\end{align}
It is easy to see that the feasible region of problem \eqref{B2} is a subset of the feasible region of problem \eqref{B1} and $({L^{\rm{*}}},{Q^{\rm{*}}})$ is optimal to both \eqref{B1} and \eqref{B2}. Thus any optimal solution of problem \eqref{B2}  must be an optimal solution of problem \eqref{B1}. Therefore, to justify that the  encoder-decoder pairs with joint distribution matrices $({{\bf{L}}^{\rm{*}}},{{\bf{L}}^{\rm{*}}})$ and $({{\bf{Q}}^{\rm{*}}},{{\bf{Q}}^{\rm{*}}})$ are optimal to \eqref{B1}, it is enough to justify that $({{\bf{L}}^{\rm{*}}},{{\bf{L}}^{\rm{*}}})$ and $({{\bf{Q}}^{\rm{*}}},{{\bf{Q}}^{\rm{*}}})$ are optimal to \eqref{B2}.

Obviously, the constraint ${p_Z}(z_j) = {h_j}$ in \eqref{B2} can be expressed as $\sum\limits_i {{l_{i,j}}}  = \sum\limits_i {{q_{i,j}}}  = {h_j}$. Besides, since $Y$ and $\hat Y$ have the same distribution under perfect perception constraint, the constraint $d({p_Y},{p_{\hat Y}}) \le 0$ can be expressed as $\sum\limits_j {{l_{i,j}}}  = \sum\limits_j {{q_{i,j}}}  = {p_Y}(y_i)$. Now, we rewrite the objective function of \eqref{B2} as
\begin{align}
\begin{split}
\frac{{\rm{1}}}{t}\mathbb{E}{\rm{[}}{\left\| {Y - \hat Y} \right\|^2}&=\frac{1}{t}\sum\limits_{y,\hat y} {{p_{Y,\hat Y}}(y,\hat y){{\left\| {y - \hat y} \right\|}^2}}\\
&=\frac{1}{t}[\sum\limits_y {{p_Y}(y){y^T}y}  + \sum\limits_{\hat y} {{p_{\hat Y}}(\hat y){{\hat y}^T}\hat y}  - 2\sum\limits_{y,\hat y} {{p_{Y,\hat Y}}(y,\hat y){y^T}\hat y} ],
\end{split}
\end{align}
where $\sum\limits_y {{p_Y}(y){y^T}y}$ is constant for fixed source, and $\sum\limits_{\hat y} {{p_{\hat Y}}(\hat y){{\hat y}^T}\hat y}  = \sum\limits_y {{p_Y}(y){y^T}y}$ for the perfect perception constraint. Hence, minimizing the objective function of \eqref{B2} is to equivalent to maximizing $\sum\limits_{y,\hat y} {{p_{Y,\hat Y}}(y,\hat y){y^T}\hat y}$, for which we have
\begin{align}
\begin{split}
\sum\limits_{y,\hat y} {{p_{Y,\hat Y}}(y,\hat y){y^T}\hat y} &=\sum\limits_{y,\hat y,z} {{p_{Y,\hat Y,Z}}(y,\hat y,z){y^T}\hat y}\\
&\mathop {\rm{ = }}\limits^{(a)} \sum\limits_{y,\hat y,z} {{p_Z}(z){p_{Y|Z}}(y|z){p_{\hat Y|Z}}(\hat y|z){y^T}\hat y}\\
&=\sum\limits_j {{h_j}[\sum\limits_i {{p_{Y|Z}}({y_i}|z_j^{}){y_i}^T\sum\limits_k {{p_{\hat Y|Z}}({{\hat y}_k}|z_j^{}){{\hat y}_k}} } ]}\\
&=\sum\limits_j {{h_j}\mathbb{E}{{(Y|Z = z_j^{})}^T}\mathbb{E}(\hat Y|Z = z_j^{})}
\end{split}
\end{align}
where in (a) we used the property of Markov chain $Y \to Z \to \hat Y$ that $Y$ and $\hat Y$ are independent under condition $Z$. Hence, using the joint distribution representation $({\bf{L}},{\bf{Q}})$ of the encoder-decoder pair $(L,Q)$, the problem \eqref{B2} can be equivalently  reformulated as
\begin{align}
\begin{split}
\mathop {\max }\limits_{{\bf{L}},{\bf{Q}}}\ &\sum\limits_j {{h_j}\mathbb{E}{{(Y|Z = z_j)}^T}\mathbb{E}(\hat Y|Z = z_j)}\\
s.t.\quad&\sum\limits_i {{l_{i,j}}}  = \sum\limits_i {{q_{i,j}}}  = {h_j},{\rm{ }}1 \le j \le n\\
&\sum\limits_j {{l_{i,j}}}  = \sum\limits_j {{q_{i,j}}}  = {p_Y}({y_i}),{\rm{ }}1 \le i \le m\\
&{\rm{0}} \le {\rm{ }}{l_{i,j}},{q_{i,j}} \le 1,{\rm{ }}1 \le i \le m,{\rm{ }}1 \le j \le n. \label{DR0function3}
\end{split}
\end{align}
Accordingly, the joint distribution matrix pair $({{\bf{L}}^{\rm{*}}},{{\bf{Q}}^{\rm{*}}})$ corresponding to the optimal encoder-decoder pair $(L^*,Q^*)$ is an optimal solution to \eqref{DR0function3}. Recall that ${\bf{L}}$ is the probability matrix of $p_{Y,Z}$ and  ${\bf{Q}}$ is the probability matrix of $p_{{\hat Y},Z}$, hence $\mathbb{E}(Y|Z = z_j^{})$ and $\mathbb{E}({\hat Y}|Z = z_j^{})$ are functions of ${\bf{L}}$ and ${\bf{Q}}$. Define
\begin{align}
&{f_j}({\bf{L}}): = \mathbb{E}(Y|Z = z_j) = \sum\limits_i {\frac{{{l_{i,j}}}}{{{h_j}}}{y_i}},\\
&{f_j}({\bf{Q}}): = \mathbb{E}(Y|Z = z_j) = \sum\limits_i {\frac{{{q_{i,j}}}}{{{h_j}}}{y_i}},
\end{align}
and
\begin{align}
\begin{split}
F({\bf{L}},{\bf{Q}}): &= \mathop \sum \limits_j {h_j}\mathbb{E}\left( {Y|Z = z_j} \right)^T\mathbb{E}\left( {\hat Y|Z = z_j} \right)\\
&= \mathop \sum \limits_j {h_j}{f_j}{({\bf{L}})^T}{f_j}({\bf{Q}}).
\end{split}
\end{align}
Next, we show $({{\bf{L}}^{\rm{*}}},{{\bf{L}}^{\rm{*}}})$ and $({{\bf{Q}}^{\rm{*}}},{{\bf{Q}}^{\rm{*}}})$ are also optimal solutions to \eqref{DR0function3}.

Because $({{\bf{L}}^{\rm{*}}},{{\bf{Q}}^{\rm{*}}})$ is an optimal solution to \eqref{DR0function3}, the optimal objective value of \eqref{DR0function3} is $F({{\bf{L}}^*},{{\bf{Q}}^*})$. Since the constraints of $\bf{L}$ and $\bf{Q}$ are the same, it is easy to see that $({{\bf{L}}^{\rm{*}}},{{\bf{L}}^{\rm{*}}})$ and $({{\bf{Q}}^{\rm{*}}},{{\bf{Q}}^{\rm{*}}})$ are both feasible solutions to \eqref{DR0function3}. Meanwhile, we have
\begin{align}
&F({{\bf{L}}^*},{{\bf{L}}^*}) = \mathop \sum \limits_j {h_j}{f_j}{({{\bf{L}}^*})^T}{f_j}({{\bf{L}}^*}) = \mathop \sum \limits_j {h_j}{\left\| {{f_j}({{\bf{L}}^*})} \right\|^2}, \label{F(L,L)}\\
&F({{\bf{Q}}^*},{{\bf{Q}}^*}) = \mathop \sum \limits_j {h_j}{f_j}{({{\bf{Q}}^*})^T}{f_j}({{\bf{Q}}^*}) = \mathop \sum \limits_j {h_j}{\left\| {{f_j}({{\bf{Q}}^*})} \right\|^2}. \label{F(Q,Q)}
\end{align}
Summing up \eqref{F(L,L)} and \eqref{F(Q,Q)} yields
\begin{align}
\begin{split}
F({{\bf{L}}^*},{{\bf{L}}^*}) + F({{\bf{Q}}^*},{{\bf{Q}}^*})
&= \mathop \sum \limits_j {h_j}\left( {{{\left\| {{f_j}({{\bf{L}}^*})} \right\|}^2} + {{\left\| {{f_j}({{\bf{Q}}^*})} \right\|}^2}} \right)\\
&\mathop  \ge \limits 2\mathop \sum \limits_j {h_j}\left\| {{f_j}({{\bf{L}}^*})} \right\|\left\| {{f_j}({{\bf{Q}}^*})} \right\|\\
&\mathop  \ge \limits^{(b)} 2\mathop \sum \limits_j {h_j}|{f_j}{({{\bf{L}}^*})^T}{f_j}({{\bf{Q}}^*})|\\
&\mathop  \ge \limits^{(c)} 2\mathop \sum \limits_j {h_j}{f_j}{({{\bf{L}}^*})^T}{f_j}({{\bf{Q}}^*})\\
&= 2F({{\bf{L}}^*},{{\bf{Q}}^*}), \label{F(L,Q)}
\end{split}
\end{align}
where in (b) we used the Cauchy inequality and (c) is due to the non-negativity of $h_j$. Since $({{\bf{L}}^*},{{\bf{Q}}^*})$ is an optimal solution to \eqref{DR0function3}, $F({{\bf{L}}^*},{{\bf{Q}}^*}) \ge F({\bf{L}},{\bf{Q}})$ holds for any $({\bf{L}},{\bf{Q}})$ under the constraint of \eqref{DR0function3}, which together with \eqref{F(L,Q)} implies $F({{\bf{L}}^*},{{\bf{L}}^*}) = F({{\bf{Q}}^*},{{\bf{Q}}^*}) = F({{\bf{L}}^*},{{\bf{Q}}^*})$. Therefore, $({{\bf{L}}^{\rm{*}}},{{\bf{L}}^{\rm{*}}})$ and $({{\bf{Q}}^{\rm{*}}},{{\bf{Q}}^{\rm{*}}})$ are also optimal solutions to \eqref{DR0function3}

Thus, for any source length $t$, there exist optimal solutions to \eqref{DR0function3} satisfying
\begin{align}
\begin{split}
p_{Y,Z} = p_{\hat Y,Z},\ p_{Y|Z} = p_{\hat Y|Z},
\end{split}
\end{align}
which finally results in Theorem 1.

\section{Derivation of equation (20)}

Equation (20) can be straightforwardly derived as
\begin{align}
\begin{split}
\frac{1}{t}\mathbb{E}{\rm{[}}{\left\| {Y - \hat Y} \right\|^2}&=\sum\limits_{y,\hat y} {{p_{Y,\hat Y}}(y,\hat y){{\left\| {y - \hat y} \right\|}^2}}\\
&\mathop  = \limits^{(d)} \frac{1}{t}\sum\limits_{y,\hat y,z} {{p_{Y,\hat Y,Z}}(y,\hat y,z){{\left\| {y - \mathbb{E}{\rm{[}}Y|z{\rm{] + }}\mathbb{E}{\rm{[}}\hat Y|z{\rm{]}} - \hat y} \right\|}^2}}\\
&\mathop  = \limits^{(e)} \frac{1}{t}\sum\limits_{y,z} {{p_{Y,Z}}(y,z){{\left\| {y - \mathbb{E}{\rm{[}}Y|z{\rm{]}}} \right\|}^2}}  + \frac{1}{t}\sum\limits_{\hat y,z} {{p_{\hat Y,Z}}(\hat y,z){{\left\| {\mathbb{E}{\rm{[}}\hat Y|z{\rm{]}} - \hat y} \right\|}^2}}\\
&\mathop  = \limits^{(f)} \frac{{\rm{2}}}{t}\mathbb{E}\left[ {{{\left\| {Y - \mathbb{E}[Y|Z]} \right\|}^2}|Z} \right],
\end{split}
\end{align}
where (d) is due to $\mathbb{E}[Y|Z]=\mathbb{E}[\hat Y|Z]$, (e) is due to
\begin{align}
\begin{split}
\sum\limits_y {{p_{Y|Z}}(y|z)(y - \mathbb{E}[Y|Z])} &= \sum\limits_y {{p_{Y|Z}}(y|z)y}  - \mathbb{E}[Y|Z]\\
&= \mathbb{E}[Y|Z]-\mathbb{E}[Y|Z] = 0
\end{split}\\
\begin{split}
\sum\limits_{\hat  y} {{p_{\hat Y|Z}}(\hat y|z)(\hat y - \mathbb{E}[\hat Y|Z])} &= \sum\limits_{\hat y} {{p_{\hat Y|Z}}(\hat y|z)\hat y}  - \mathbb{E}[\hat Y|Z]\\
&= \mathbb{E}[\hat Y|Z]-\mathbb{E}[\hat Y|Z] = 0
\end{split}
\end{align}
and (f) is due to the same distribution of $Y$ and $\hat Y$.

\section{Degenerate problem}

\begin{figure*}[!t]
	\vskip 0.2in
	\begin{center}
		\subfigure[MSE loss versus training epoch]{
			\begin{minipage}[t]{0.45\columnwidth}
				\centering
				\includegraphics[width=0.9\columnwidth]{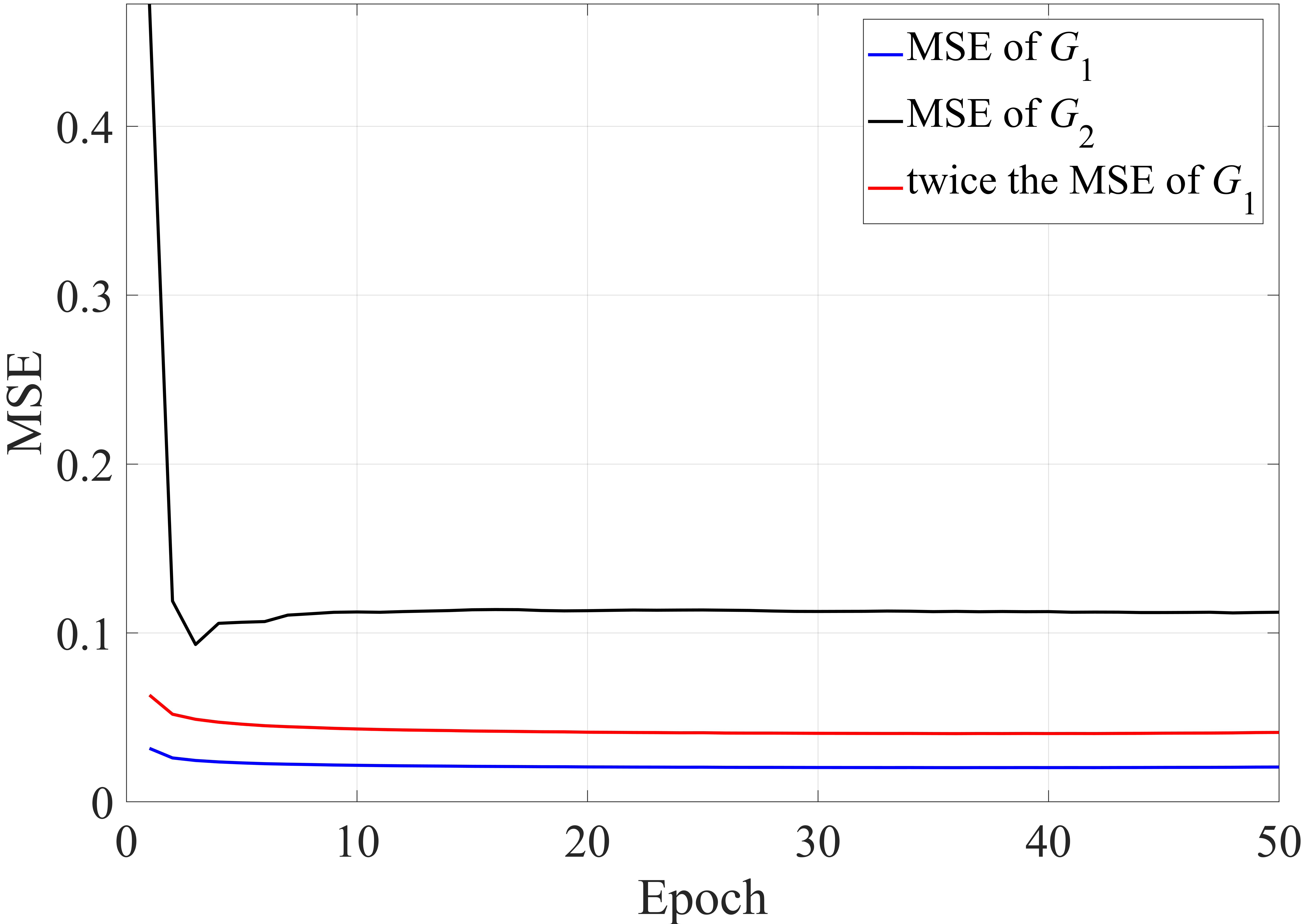}
			\end{minipage}%
		}%
		\subfigure[Visual comparison of the output]{
			\begin{minipage}[t]{0.45\columnwidth}
				\centering
				\includegraphics[width=0.9\columnwidth]{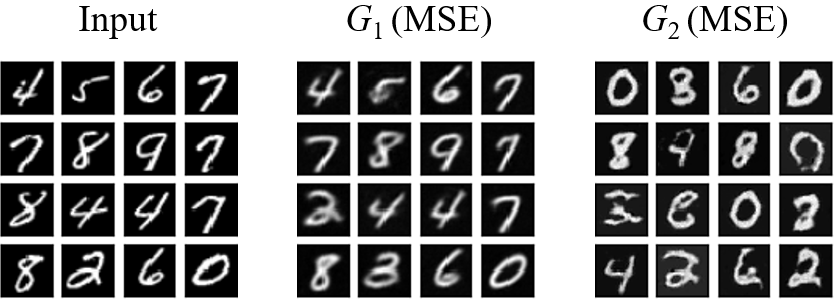}
			\end{minipage}%
		}
		\caption{\textbf{Illustration of a typical degeneration case.}}
		\label{Degradation}
	\end{center}
	\vskip -0.2in
\end{figure*}

Figure~\ref{Degradation} shows a degeneration case in training $G_2$, where the MSE of $G_2$ converges to a value deviates largely from the 2-fold MSE of $G_1$. From Fig. 7(b), while the output numbers of $G_1$ are correct, those of $G_2$ are incorrect though more clear. It means that the bit stream from $E$ contains enough information for correctly reconstructing the numbers, but the trained model $G_2$ tends to generate numbers randomly. This problem is typically encountered in adversarial training, due to that the the alternating training procedure converges to a poor point. To address this problem, we pre-train the discriminator $J$ to discriminate between $({x_i},E({x_i}))$ and $({x_j},E({x_i}))$ with $i \ne j$, where $x_i$ and $x_j$ are samples of $X$. Intensive experiments show that this strategy can effectively reduce the occurrence of the degeneration problem.

\end{document}